\documentclass[conference]{IEEEtran}
\usepackage{graphicx}
\usepackage{amsmath}
\usepackage{booktabs}
\usepackage{url}
\graphicspath{{figures/}}

\title{How Stable Is a PNT Resilience Score? Decision-Instability of\\ Single-Number Resilience Ratings under Framework-Aligned Weighting}

\author{\IEEEauthorblockN{Chakshu Baweja}
\IEEEauthorblockA{Ashforde O\"U\\
Tallinn, Estonia\\
contact@ashforde.org}}

\begin{document}
\maketitle

\begin{abstract}
Every authoritative positioning, navigation, and timing (PNT) resilience framework, from the DHS Resilient PNT Conformance Framework (RPCF) to the Resist-Detect-Respond-Recover model and Yang's resilient-PNT criteria, defines what resilience means but supplies only self-attestation: a checklist or a maturity Level, with no engine, no measurement, and no evidence. We build the missing measurement layer as an open, deterministic scoring engine over a PNT simulator, emitting per-dimension sub-scores each traceable to a scenario and an oracle and each tagged with its honest validation status. We then ask whether a single composite score, or a single maturity Level, is a stable basis for a decision. Across a reference panel of seven architectures built to span genuine cross-dimension tradeoffs, a Dirichlet weighting simplex over the seven RPCF categories, and a five-threat ensemble, the answer separates into two regimes. The composite winner is stable under active denial and under near-equal weightings, flipping in about 1 percent of draws, so a single number is safe precisely where one design dominates; but re-weighting alone flips the winner in up to 22 percent of draws under nominal conditions, where designs genuinely contend, and that instability is a domain application of known composite-indicator sensitivity. The sharper and weighting-invariant failure is categorical: a weakest-link maturity Level (our minimum-over-categories operationalization of the RPCF ladder, not the framework's own rule) is a function of the threat assumed rather than of the architecture, changing for one architecture in seven across the ensemble. A constructed example shows that, because the composite rewards declared techniques, a single-band receiver declaring all seven techniques can outscore a genuinely more resilient system: self-attestation can be gamed by declaration. And apparent fourfold GNSS redundancy reduces, by the definition of a shared common-mode failure domain, to an effective diversity of one. The conclusions hold under a plus-or-minus 20 percent perturbation of every driver within the modelled reduction. We argue for reporting per-dimension sub-scores with provenance and a rank range, not a phantom single number. This is a simulation-derived self-assessment aligned to RPCF v2.0, not a certification.
\end{abstract}

\section{Introduction}
A recurring gap runs through the PNT-resilience literature. Resilient PNT is a recognised national priority \cite{eo13905}, and government, industry, and academic sources converge on the need for a quantified, reproducible, vendor-neutral resilience score checked against a published framework, yet every fielded instrument is self-attestation or pure concept. The DHS Resilient PNT Conformance Framework (RPCF) v2.0 \cite{rpcf} defines maturity Levels 0 to 4 and seven technique categories (Obfuscate, Limit, Verify, Isolate, Diversify, Mitigate, Recover). The Resist-Detect-Respond-Recover (RDRR) function model \cite{rethinkpnt} frames an assurance case as a documented body of evidence. Yang's resilient-PNT concept \cite{yang} names availability, reliability, continuity, and accuracy criteria. None ships a measurement: a system owner declares conformance, or an assessor scores a questionnaire, and the result is a number nobody can reproduce.

This invites a hazard that is rarely examined. To produce a single grade, the multi-dimensional, model-derived reality of a PNT system must be collapsed into one number or one Level. Composite-indicator methodology has long warned that such collapses are sensitive to the (arbitrary) weighting and aggregation choices \cite{saisana,oecd,saltelli}, but the warning has not been quantified for PNT resilience, where the dimensions trade off in physically meaningful ways: a single-band receiver is precise and cheap but fragile, while a diverse architecture is robust but each fallback source is coarser.

We make three contributions. First, an open, deterministic scoring engine that maps a PNT architecture and its simulated behaviour to per-dimension sub-scores across the RPCF categories, the RDRR functions, and Yang's criteria, with every sub-score traceable to a scenario, a seed, an engine version, and an oracle, and tagged with its validated-or-modelled status. Second, a study of the decision-stability of the resulting single score and single Level, under a defensible weighting space and a threat ensemble. Third, a reporting protocol that follows from the result.

The decision-stability question has a direct precedent in composite-indicator sensitivity analysis \cite{saisana,oecd,saltelli}; our weighting-instability result (Section~\ref{sec:results}) is that established method applied to a new domain, and we claim it as such, not as a new technique. The genuinely framework-specific finding is categorical and lies outside weighted-aggregate sensitivity: a weakest-link Level, a minimum over the categories (our operationalization of the RPCF ladder, not the framework's own definition), is driven by the threat assumed and is invariant to the weighting. Two further points are domain properties rather than sensitivity results: because the composite rewards declared techniques independently of measured behaviour, self-attestation can be gamed by declaration, which we show by construction; and an apparently redundant architecture whose sources share a common-mode failure domain has, by definition, an effective diversity far below its source count.

\textbf{Honest scope.} This is a controlled, parameter-grounded simulation, not a field measurement, and it is a simulation-derived self-assessment aligned to RPCF v2.0, not a certification, accreditation, or compliance statement, and not endorsed by any authority. The sub-scores are timing-domain and detection figures of merit; they are not position-domain accuracy. The modelled-majority of the underlying engine is carried on the face of every result.

\section{Frameworks and related work}
\textbf{PNT-resilience frameworks.} RPCF v2.0 \cite{rpcf} is the most operationally specific: it ladders systems through Levels 0 to 4 against the seven technique categories above, and is moving toward formal standardization (IEEE P1952 \cite{p1952}). The RDRR model \cite{rethinkpnt} and Yang's framework \cite{yang} are complementary lenses on the same behaviours. All three are declarative; the present work supplies a measurement layer aligned to them, never replacing or certifying against them.

\textbf{Resilience quantification.} The notion of a resilience curve, the loss and recovery of function over time, originates in the resilience-triangle of Bruneau et al. \cite{bruneau}; our timeline metrics (detection, reaction, recovery, loss duration) and the bounded-versus-unbounded degradation verdict are PNT instances of that idea.

\textbf{Composite-indicator fragility.} That a weighted aggregate can reorder the things it ranks is known in the composite-indicator literature \cite{saisana,oecd}; our contribution is to measure the effect for PNT resilience and to tie it to the declared-versus-measured gap that self-attestation creates. We quantify ranking agreement with Kendall's tau \cite{kendall}.

\section{Method}
\subsection{Architecture model}
An architecture is a named set of PNT sources, each with a kind, an independence group (sources sharing a group fail together), and a quality weight, plus the set of RPCF technique categories it declares. Independence groups separate apparent redundancy (source count) from effective redundancy (independent failure domains).

\subsection{Framework-aligned sub-scores}
Each (architecture, scenario) pair is reduced to a behaviour summary: holdover (in-spec coast under denial), availability, impairment-detector AUC, integrity, and a bounded-degradation flag. Each RPCF category sub-score is a documented, monotone function of one or more of these drivers in $[0,1]$: Verify from detector AUC; Diversify from the inverse-Simpson effective diversity (the Hill number of order 2) over independence groups \cite{simpson,hill}; Mitigate from availability; Recover from holdover gated by the bounded flag; and the procedural categories (Obfuscate, Limit, Isolate) from declared presence scaled by source quality, so these three are declaration-driven and scenario-independent, which is precisely the mechanism H2 exploits. The RDRR functions and Yang criteria are documented re-projections of the same drivers and add no independent information; the position-domain Yang accuracy criterion is not modelled and is carried as a gap. Every sub-score carries a validated-or-modelled status and an oracle kind; in this study all drivers are modelled, and the engine's machine-checked verification matrix forbids labelling a self-consistency check as an external validation.

\subsection{Composite and tentative Level}
The composite is a weight-normalized mean of the seven RPCF sub-scores, the single number whose stability we interrogate. We also assign a maturity Level by our own operationalization of the RPCF ladder, which RPCF v2.0 specifies qualitatively rather than as a formula: a weakest-link ladder on the minimum sub-score (cutpoints at $0.2/0.4/0.6/0.8$), with a bounded-degradation gate that caps maturity at Level 2 when error is unbounded, on the reasoning that a system that cannot bound its error under sustained denial cannot substantiate a higher Level. The minimum rule, the cutpoints, and the gate are our design choices, not RPCF's definition; we call the result a \emph{tentative} Level throughout, and the threat-dependence we report is a property of this weakest-link operationalization.

\subsection{Instability metrics}
We sample dimension weightings from a symmetric Dirichlet over the seven RPCF categories. At concentration $\alpha=1$ this is uniform on the simplex (maximum entropy over weight vectors); it deliberately includes near-degenerate weightings that load most of the mass on a single category, so it is a \emph{broad} space rather than a curated set of expert-endorsed weightings, and the concentration sweep below tightens it toward equal weights. For each (weighting, scenario) draw we score and rank every architecture, and report the top-1 flip rate (the fraction of draws whose winner differs from the modal winner), which we give \emph{per scenario}, because pooling across scenarios mixes a high-instability nominal regime with near-stable active-denial regimes and a single pooled rate would be a mixture of heterogeneous blocks rather than one proportion. We also report the mean Kendall tau \cite{kendall} between each draw's ranking and its scenario's equal-weight reference, the Level flip rate (the fraction of architectures whose tentative Level is not constant across scenarios), and each architecture's rank range. The equal-weight ranking is a descriptive reference, not a privileged truth, since the thesis is that no weighting is privileged; Kendall tau measures whole-order agreement and quantizes in steps of $2/\binom{7}{2}\approx0.095$ for seven items, so it understates instability concentrated in specific middle ranks and we read it coarsely alongside the rank ranges. To separate the two sources of variation we report the weighting-only flip rate within each fixed scenario (re-weighting with the threat held constant) and sweep the concentration $\alpha\in\{1,2,5,10,20\}$ from broad to near-equal. Finally, to probe robustness to the chosen driver values, we perturb every behaviour driver by a uniform $\pm 20$ percent over 40 replicates and recompute the metrics; this varies driver magnitudes but holds the functional form and the decision thresholds fixed, so it tests robustness within the reduction, not of the reduction itself.

\subsection{Reference panel and threat ensemble}
The panel has seven architectures spanning single-band GNSS, multi-band GNSS, GNSS with inertial, a spoof-hardened variant, a deliberately apparent-redundant quad-GNSS array, a ``paper-tiger'' single-band receiver that nonetheless declares all seven techniques, and a genuinely diverse architecture (GNSS, inertial, clock, eLoran). The threat ensemble is nominal, wideband jamming, spoofing, meaconing, and a combined attack, the standard civil GNSS threat classes \cite{psiaki}. The reduction from (architecture, threat) to a behaviour summary is a documented, deliberately simple physical model: a denial defeats GNSS RF sources; surviving non-GNSS sources set holdover; detection depends on the declared Verify technique and on cross-checks from independent groups; and a stealthy spoof corrupts the solution silently when the modelled detector AUC fails to clear a fixed threshold. The panel deliberately includes adversarial constructions, the paper-tiger receiver for H2 and the apparent-redundant array for H3, to surface specific failure modes; we therefore do not claim the panel is unbiased, only that, given the panel, the instability metrics are computed without further tuning. The bounded-degradation verdicts depend on where the modelled AUC values sit relative to the detection threshold, a sensitivity we flag in the limitations.

\subsection{Pre-registered hypotheses}
\textbf{H1a (weighting instability).} Within a fixed threat, re-weighting reorders the top architecture. Refuted if the winner is stable in at least 95 percent of weight draws under a near-equal (defensible) prior. \textbf{H1b (Level threat-dependence).} The tentative weakest-link Level is a function of the threat assumed, not a property of the architecture. Refuted if every architecture holds a constant Level across the threat ensemble. \textbf{H2 (measured beats declared).} Architectures with identical declared techniques can have opposite measured bounded verdicts, and Levels, under the same threat; established by a constructed existence example, not a survey statistic. \textbf{H3 (apparent vs effective redundancy).} Under a shared common-mode failure domain, an apparently redundant architecture's effective diversity is far below its source count, a consequence of the failure-domain assignment that the engine deliberately separates from raw source count.

\section{Results}
\label{sec:results}
The study evaluates seven architectures across five threats and 2000 weight draws per scenario, that is 10{,}000 rankings (and 70{,}000 architecture evaluations) per weighting prior, engine version 0.19.0, fixed seed. Table~\ref{tab:summary} consolidates the per-architecture picture.

\begin{table}[t]
\caption{Per-architecture summary. Composite is the nominal-scenario equal-weight score; Level range and rank range are over the five threats and the broad weighting simplex; diversity is effective (inverse-Simpson) assuming independence then accounting for GNSS common-mode.}
\label{tab:summary}
\centering
\footnotesize
\begin{tabular}{lcccc}
\toprule
Architecture & Comp. & Level & Rank & Diversity \\
\midrule
diverse\_full   & 0.80 & 0--3 & 1--2 & 3.9$\rightarrow$3.9 \\
checkbox\_gnss  & 0.69 & 0    & 1--5 & 1.0$\rightarrow$1.0 \\
spoof\_hardened & 0.45 & 0    & 2--4 & 1.9$\rightarrow$1.9 \\
quad\_gnss      & 0.45 & 0    & 2--5 & 4.0$\rightarrow$1.0 \\
gnss\_ins       & 0.34 & 0    & 3--5 & 1.9$\rightarrow$1.9 \\
gnss\_multiband & 0.29 & 0    & 6--7 & 1.0$\rightarrow$1.0 \\
gnss\_l1        & 0.29 & 0    & 6--7 & 1.0$\rightarrow$1.0 \\
\bottomrule
\end{tabular}
\end{table}

\subsection{H1a: weighting instability is confined to the contested middle}
\begin{figure}[t]
\centering
\includegraphics[width=\columnwidth]{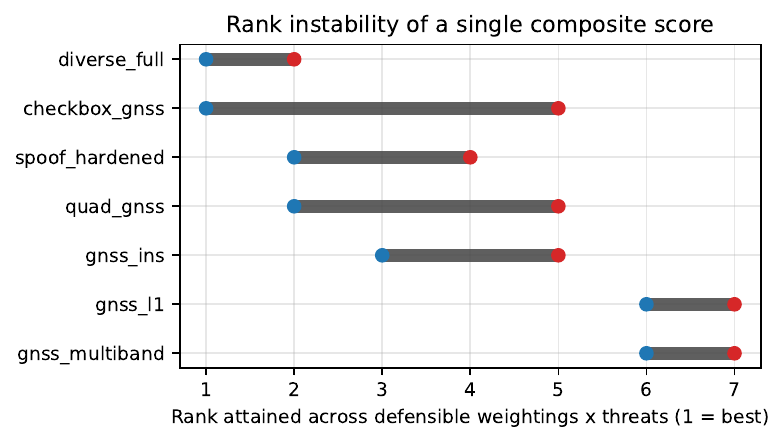}
\caption{Rank attained by each architecture across defensible weightings and threats (1 = best). The genuinely diverse design is stable at the top and the bare receivers at the bottom; the contested middle, including a checklist-compliant ``paper tiger,'' swings across most of the ranking.}
\label{fig:ranks}
\end{figure}

\begin{figure}[t]
\centering
\includegraphics[width=\columnwidth]{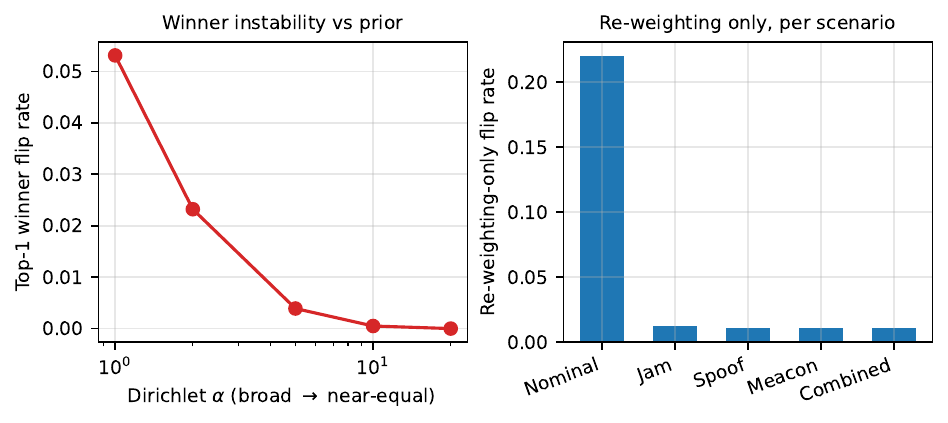}
\caption{Left: the pooled top-1 winner flip rate falls monotonically as the Dirichlet concentration $\alpha$ tightens from broad ($\alpha=1$) toward near-equal ($\alpha=20$), so the instability is a property of broad, defensible weightings rather than of extremes. Right: separating the two sources of variation, re-weighting alone with the threat held fixed flips the winner in 22 percent of draws under nominal conditions but in about 1 percent under each active-denial scenario.}
\label{fig:weight}
\end{figure}

The winner-stability result splits cleanly by prior (Fig.~\ref{fig:ranks}, Fig.~\ref{fig:weight}). Pooled across the five threats, the modal winner is chosen in 94.7 percent of broad-simplex ($\alpha=1$) draws (top-1 flip rate 0.053) and in 99.6 percent under the near-equal prior (0.004). \emph{Under the near-equal, defensible prior H1a is refuted}: the winner clears the 95 percent stability bar comfortably. It survives only under the broad uniform-simplex prior, which narrowly fails the bar (94.7 percent) and does so largely because that prior admits near-degenerate weightings no analyst would choose. Weighting instability is therefore not this paper's strong claim; the categorical Level instability of Section~\ref{sec:levelflip} is.

Where weighting instability does live is worth locating, because it is the case a buyer cares about. Re-weighting alone, with the threat held fixed, flips the winner in 22.0 percent of draws under nominal conditions but in only about 1 percent under each active-denial scenario: when GNSS is denied the genuinely diverse architecture dominates regardless of weighting, but when nothing is denied a checklist-declaring single-band receiver competes with it and the winner depends on which dimensions one weights. This is the weighted-aggregate sensitivity that the composite-indicator literature \cite{saisana,oecd,saltelli} describes, in a PNT instance. The rank ranges of the non-dominant architectures span most of the order (Fig.~\ref{fig:ranks}): the paper-tiger single-band receiver ranges over ranks one to five, the apparent-redundant quad-GNSS array two to five, and the GNSS-inertial system three to five, while only the dominant architecture (ranks one to two) and the bare receivers (ranks six to seven) are stable. The mean Kendall tau to the equal-weight reference is 0.90 under the broad prior, which confirms that the aggregate order is largely stable and that the instability is concentrated in the middle ranks rather than spread across the ranking.

Two controls frame the reading. Sweeping the Dirichlet concentration, the pooled flip rate falls monotonically from 0.053 at $\alpha=1$ to near zero at $\alpha=20$ (Fig.~\ref{fig:weight}), so what instability there is comes from broad priors, not from any single extreme. And under a $\pm 20$ percent perturbation of every driver across 40 replicates, the contested-middle architectures keep wide rank ranges in 100 percent of replicates and the Level-flip rate is invariant, so the conclusions are not knife-edge in the driver values; robustness to the functional form and the decision thresholds is a separate question we address in the limitations.

\subsection{H1b: a single Level is scenario-dependent}
\label{sec:levelflip}
\begin{figure}[t]
\centering
\includegraphics[width=\columnwidth]{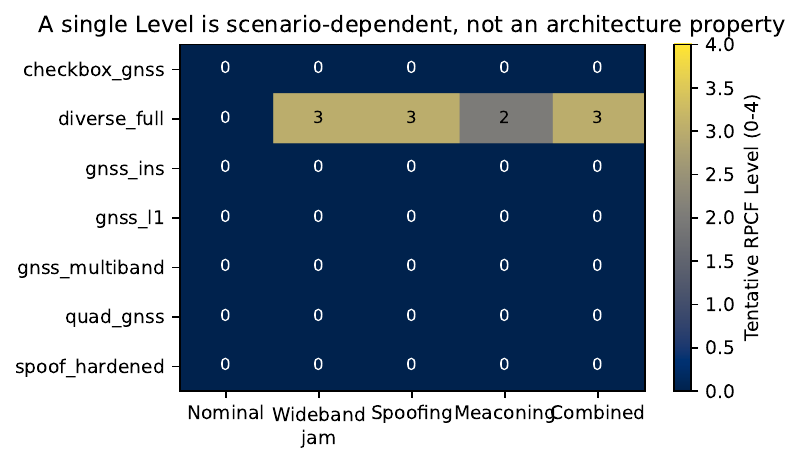}
\caption{Tentative RPCF Level by architecture and threat. The same diverse architecture is graded Level 0 to 3 depending on the scenario assumed, while a single declared Level would report one number.}
\label{fig:level}
\end{figure}

This is the paper's strongest and most weighting-invariant finding. The tentative weakest-link Level (our minimum-over-categories operationalization of the RPCF ladder, not RPCF's own rule) is not a property of the architecture but of the threat assumed (Fig.~\ref{fig:level}). The diverse architecture is rated Level 3 under jamming, spoofing, and the combined attack, Level 2 under meaconing (harder to detect), and Level 0 under nominal conditions, where no denial occurs to demonstrate recovery. One in seven architectures changes Level across the ensemble (14.3 percent), so H1b holds. Unlike the weighting instability above, this is invariant to the weighting and is not reachable by weighted-aggregate sensitivity analysis, because a minimum ladder, not a mean, drives it: a single Level label hides the threat-dependence entirely.

\subsection{H2: declared posture is not measured resilience}
\begin{figure}[t]
\centering
\includegraphics[width=\columnwidth]{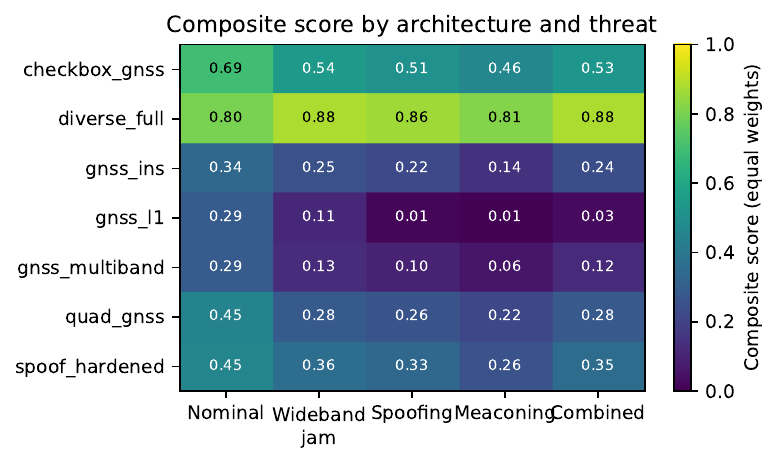}
\caption{Composite score (equal weights) by architecture and threat. A checklist-compliant single-band receiver outscores a genuinely more resilient GNSS-inertial system purely by declaring techniques.}
\label{fig:composite}
\end{figure}

The composite rewards declared techniques: the paper-tiger single-band receiver, which declares all seven RPCF categories, scores 0.69 under equal weights in nominal conditions, second of seven and ahead of a real GNSS-inertial system at 0.34 (Fig.~\ref{fig:composite}). Yet a checklist sees the paper-tiger and the diverse architecture as identical (both declare all seven techniques), while their measured behaviour diverges completely: across all four denial scenarios the paper-tiger is unbounded and Level 0, the diverse architecture is bounded and Level 2 to 3 (four declared-versus-measured contrasts). We construct this pair deliberately, so it is an existence proof rather than a survey statistic: it shows that because the composite and the technique checklist reward declared categories, a posture that is identical on paper can hide opposite measured resilience. Self-attestation can therefore be gamed by declaration; this single constructed pair establishes existence, not prevalence.

\subsection{H3: apparent redundancy is illusory}
\begin{figure}[t]
\centering
\includegraphics[width=\columnwidth]{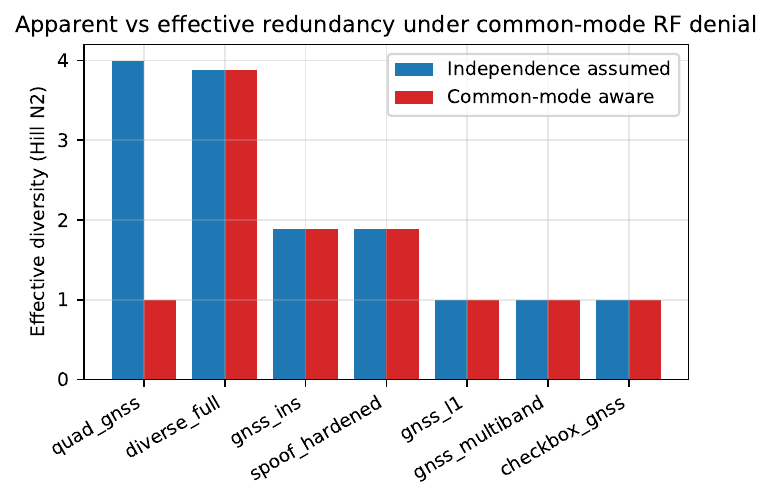}
\caption{Effective diversity assuming independence versus accounting for GNSS common-mode coupling. Apparent fourfold redundancy collapses to one; genuine cross-domain diversity survives.}
\label{fig:diversity}
\end{figure}

A four-receiver GNSS array in four independence groups has an effective diversity (inverse-Simpson, the Hill number of order 2) of approximately 4.0 (3.998 with the modelled quality weights) when the groups are assumed independent. Once all four are assigned to a single GNSS common-mode failure domain, as a wideband denial that defeats them at once requires, that number is 1.0 by the definition of the index (Fig.~\ref{fig:diversity}). The collapse is therefore a consequence of the failure-domain assignment, not an emergent measurement; the substantive and externally well-established content is the physical premise that co-located GNSS receivers share an RF failure domain. The genuinely diverse architecture, whose clock, inertial, and eLoran sources do not share that domain, retains an effective diversity of 3.9. The point for scoring is that counting sources, or even counting nominal independence groups, overstates redundancy a common-mode threat erases, and that the engine separates the two so the overstatement is visible. Inverse-Simpson over quality-weighted groups is a redundancy proxy and carries no failure-probability semantics; a reliability-weighted survival metric is future work.

\section{Discussion}
The results argue for a specific reporting discipline. Do not report a single composite resilience score or a single maturity Level. Report the per-dimension sub-scores, each with its validated-or-modelled provenance; report the rank range an architecture occupies across defensible weightings, not a point rank; and report the assigned Level per threat scenario, not as a context-free badge. The single number is safe only when one design dominates, which is exactly the case in which the buyer does not need it; in the contested case where the buyer does need it, the number is unstable.

This complements, rather than competes with, RPCF and IEEE P1952. A standard defines the dimensions and the maturity ladder; an open, oracle-checked engine \cite{kshana} can supply reproducible per-dimension evidence and an honest account of its own uncertainty. The honesty discipline, the validated-versus-modelled split carried on the face of every report and the refusal to claim certification, is not a disclaimer bolted onto a product; for a skeptical insurer or procurement buyer it is the product.

\section{Limitations}
The architectures and threat responses are synthetic, parameter-grounded reductions, not field measurements; the instability we measure is a property of how the dimensions combine, and the qualitative conclusions depend on architectures genuinely trading off across dimensions, which real PNT systems do, rather than on the specific reduction. The figures of merit are timing-domain and detection metrics; position-domain accuracy is not modelled and is reported as a gap, not a number. The sub-score mappings are one defensible choice among several; the study's point is precisely that the choice matters, so we publish the mappings, the seeds, and the engine version for inspection. We tested robustness to a $\pm 20$ percent perturbation of every driver and the qualitative conclusions held in all 40 replicates, but that perturbation holds the functional form and the decision thresholds (the detection-AUC cutoff and the Level cutpoints) fixed; the bounded-degradation and Level verdicts depend on where the modelled detector AUC values fall relative to that detection threshold, and robustness to the thresholds and cutpoints themselves is untested. The mappings remain one defensible family, and a fundamentally different reduction could behave differently. The engine is modelled-majority, and the scores inherit that status.

\section{Reproducibility}
The scoring engine, the instability study, the reference panel and threat ensemble, the integrity-hashed assurance report, and the artifact generator are open source in the Kshana PNT-resilience simulator \cite{kshana}. Every numeric routine carries a hand-derived oracle test (35 tests for the resilience module; the full engine suite passes). The study artifact is regenerated by one command and records the engine version, seeds, weighting priors, and a configuration hash; the example assurance report carries a SHA-256 integrity hash.

\section{Conclusion}
A single-number PNT resilience score reliably identifies a dominant architecture and is stable under defensible near-equal weightings, but it reorders contesting designs under broad weightings, and, more sharply, a tentative weakest-link maturity Level is a function of the threat assumed rather than of the system. A constructed example shows declared techniques can diverge from measured resilience, and an apparently redundant architecture's effective diversity falls far below its source count once a common-mode failure domain is acknowledged. The remedy is not a better single number but the discipline of reporting per-dimension, provenance-tagged sub-scores with their ranges, which an open, oracle-checked engine can supply as a measurement layer aligned to the published frameworks.


\begin{thebibliography}{1}
\bibitem{rpcf} U.S. Department of Homeland Security, Science and Technology Directorate, ``Resilient Positioning, Navigation, and Timing (PNT) Conformance Framework,'' Version 2.0, 2022.
\bibitem{eo13905} Executive Office of the President, ``Executive Order 13905: strengthening national resilience through responsible use of positioning, navigation, and timing services,'' \emph{Federal Register}, vol.~85, no.~32, p.~9359, Feb. 2020.
\bibitem{rethinkpnt} RethinkPNT, ``A structured approach to achieving system resilience for Position, Navigation and Timing (PNT) systems,'' white paper, 2022. [Online]. Available: \url{https://rethinkpnt.com}
\bibitem{yang} Y. Yang, ``Concepts of comprehensive PNT and related key technologies,'' \emph{Acta Geodaetica et Cartographica Sinica}, vol. 45, no. 5, pp. 505--510, 2016.
\bibitem{p1952} IEEE, ``P1952: Standard for Resilient Positioning, Navigation, and Timing (PNT) User Equipment,'' standard in development.
\bibitem{psiaki} M. L. Psiaki and T. E. Humphreys, ``GNSS spoofing and detection,'' \emph{Proc. IEEE}, vol. 104, no. 6, pp. 1258--1270, 2016.
\bibitem{bruneau} M. Bruneau et al., ``A framework to quantitatively assess and enhance the seismic resilience of communities,'' \emph{Earthquake Spectra}, vol. 19, no. 4, pp. 733--752, 2003.
\bibitem{saisana} M. Saisana, A. Saltelli, and S. Tarantola, ``Uncertainty and sensitivity analysis techniques as tools for the quality assessment of composite indicators,'' \emph{J. R. Statist. Soc. A}, vol. 168, no. 2, pp. 307--323, 2005.
\bibitem{oecd} M. Nardo, M. Saisana, A. Saltelli, S. Tarantola, A. Hoffmann, and E. Giovannini, ``Handbook on Constructing Composite Indicators: Methodology and User Guide,'' OECD Publishing, 2008.
\bibitem{saltelli} A. Saltelli et al., \emph{Global Sensitivity Analysis: The Primer}. Chichester, U.K.: Wiley, 2008.
\bibitem{kendall} M. G. Kendall, ``A new measure of rank correlation,'' \emph{Biometrika}, vol. 30, no. 1/2, pp. 81--93, 1938.
\bibitem{simpson} E. H. Simpson, ``Measurement of diversity,'' \emph{Nature}, vol. 163, no. 4148, p. 688, 1949.
\bibitem{hill} M. O. Hill, ``Diversity and evenness: a unifying notation and its consequences,'' \emph{Ecology}, vol. 54, no. 2, pp. 427--432, 1973.
\bibitem{kshana} C. Baweja, ``Kshana: an open, reproducible PNT-resilience simulator,'' software, version 0.19.0, AGPL-3.0-only, 2026. [Online]. Available: \url{https://github.com/AshfordeOU/kshana}
\end{thebibliography}
\end{document}